\documentclass{ws-ijmpe}
\usepackage[super,compress]{cite}
\begin{document}

\markboth{Lizhu Chen, Yeyin Zhao, Xiaobing Li, Zhiming Li, Yuanfang Wu}{Critical properties of  various sizes  of cluster in the Ising percolation transition}

\catchline{}{}{}{}{}

\title{Critical properties of  various sizes  of cluster \\ in the Ising percolation transition}

\author{Lizhu Chen}
\address{School of Physics and Optoelectronic Engineering, \\ Nanjing University of Information Science and Technology, Nanjing 210044, China\\
chenlz@nuist.edu.cn}

\author{YeYin Zhao}
\address{Key Laboratory of Quark and Lepton Physics (MOE) and
Institute of Particle Physics, \\ Central China Normal University, Wuhan 430079, China}

\author{Xiaobing Li}
\address{Key Laboratory of Quark and Lepton Physics (MOE) and
Institute of Particle Physics, \\ Central China Normal University, Wuhan 430079, China}

\author{Zhiming Li}
\address{Key Laboratory of Quark and Lepton Physics (MOE) and
Institute of Particle Physics, \\ Central China Normal University, Wuhan 430079, China}

\author{Yuanfang Wu}
\address{Key Laboratory of Quark and Lepton Physics (MOE) and
Institute of Particle Physics, \\ Central China Normal University, Wuhan 430079, China}

\maketitle


\begin{abstract}
It is proposed that the $O(n)$ spin and geometrical percolation models can help to study the QCD phase diagram due to the universality properties of the phase transition.
In this paper,  correlations and fluctuations of various sizes of cluster in the Ising model are systematically studied.
 With a finite size system, we demonstrate how to use the finite size scaling and fixed point  behavior to search for critical point.  At critical point, the independency of system size is found from skewness and kurtosis of the maximum, second and third largest cluster and their correlations. It is similar to the Binder-ratio, which has provided a remarkable identification of the critical point.
 Through an explanation  of the universal characteristic of skewness and kurtosis of the order parameter,
a possible application to the relativistic heavy-ion collisions is also discussed.
\end{abstract}


\ccode{PACS numbers: 25.75.Gz, 25.75.Nq, 24.10.Pa}


\section{Introduction}

In high-energy nuclear collisions, we study properties of the excited nuclear matter with QCD degrees of freedom and search for the signal of the QCD phase transition.  Lattice QCD has shown that it occurs a smooth crossover from hadronic phase to the QGP phase at $\mu_B=0$ MeV~\cite{crossover}.
It is expected that the phase transition turns to be a first order at high baryon density and low temperature~\cite{first-order-1, first-order-2}. The endpoint of the first order phase transition to the crossover is referred to as the critical point, which has been studied by high-order cumulants of conserved quantities at RHIC/STAR~\cite{cumulant-proton, cumulant-charge, cumulant-kaon}.

Theoretically, Lattice QCD is of course the first principle to study the QCD phase transition~\cite{crossover, qcd-transition-Karsch-1, qcd-transition-Karsch-2, qcd-transition-Fodor, qcd-transition-Gupta}. However, simulations for much larger $\mu_B$ have sign problem~\cite{qcd-sign-problem}.  Currently, the $O(1)$ (Ising model), $O(2)$ and $O(4)$ spin models are widely applied to study the QCD phase transition due to the universality of the critical phenomena~\cite{On}, although the system is totally different.
It has been pointed out that  the  QCD critical point belongs to the same universality class as  3 dimensional (3D) Ising model.  Using a parametric representation or  Monte Carlo simulations,  the critical behavior of the high-order cumulants~\cite{stephanov-1, stephanov-2}, effects of the critical slowing down~\cite{Ising-slowing-down-1, Ising-slowing-down-2}, and the finite size scaling~\cite{finite-size-Roy-1, finite-size-Roy-2} are extracted and applied to study the QCD critical point. Especially, characteristics of the non-monotonic and sign changes in high-order cumulants are taken as the most important signals for probing the QCD critical point in experiment~\cite{STAR-non-monotonic}.

Besides the $O(n)$ spin models, the 2 dimensional (2D) geometrical percolation models can also help us to understand the QGP formation~\cite{percolation-QGP-1, percolation-xu, percolation-yu, percolation-2d, percolation-2d-2, percolation-QGP-2}.
 With growing energy and size of the colliding nuclei, the number of color strings grows and starts to overlap to form different sizes of the cluster on the transverse nucleon interaction plane, very much similar to the disks in the 2D percolation theory. Currently, the 2D percolation approach within Color String Percolation Model (CSPM) had already given a successfully description of most of the experimental data in the soft region. The CSPM can also describe the QCD to hadron cross-over transition through the temperature dependence of the shear viscosity over entropy density ratio ($\eta/s$)~\cite{ percolation-2d-2}.

It should be noted, however, the critical properties in CSPM are derived from Ref.~[\refcite{percolation-Stauffer}], which are related to the percolation transition with infinite size of the system. In heavy-ion collisions, the effect of the limited size of the formed matter can not be neglected. Consequently, we should systematically study the 2D percolation transition with finite size of the system. By a suitable definition of the cluster in the 2D Ising model, it has been found that  the thermal magnetization transition can be directly mapped into a geometrical percolation transition~\cite{Ising-temperature}.

The typical quantities in geometrical percolation transition are the percolation
strength ($P$), the maximum largest size of cluster ($\rm S_{max}$), the average
cluster size ($\left<S\right>$) and so on~\cite{percolation-cluster}. Besides these observables,
 correlation and fluctuation of the other differently ranked sizes of  cluster, such as the second largest cluster ($\rm S_{2nd}$) and the third largest cluster ($\rm S_{3rd}$), are also meaningful. We will systematically study correlation and fluctuations of the differently ranked sizes of the cluster in 2D Ising percolation transition with the finite size of the system.

Due to the finite size of the system, we should use the finite size scaling to study the critical behavior. In particular,  if a critical related observable obey the following formula of the finite size scaling~\cite{Binder}:
\begin{equation}\label{FSS-fixed-directly-1}
Q(T, L)=F_Q(tL^{1/\nu}).
\end{equation}
In this case, at $T=T_c$, we can get
\begin{equation}\label{FSS-fixed-directly-2}
Q(T_c, L)=F_Q(0).
\end{equation}

It means the temperature dependence of $Q(T, L)$ with different system sizes will intersect to a fixed point at $T = T_c$.
The critical point can be directly obtained from the temperature dependence of $Q(T, L)$. In addition, Eq.~(\ref{FSS-fixed-directly-2}) doesn't contain any critical exponent. This feature is independent of the dimensionality $D$ of the lattice or the number of dimensions of the order parameter. Consequently, it is universal both in 2D and 3D $O(n)$ models, i.e. $n=1, 2, 4$.
One most typical observable is the Binder-ratio, which has provided a remarkable identification of the critical point in simulations.
In this paper, we will show that the fixed point behavior can also be directly observed from skewness and kurtosis of the differently ranked sizes of the cluster. In addition, we will prove that the fixed point behavior for skewness and kurtosis is also universal in 2D and 3D $O(n)$ models.

This paper is organized as the following: In section 2, we will discuss the thermal magnetization transition and geometrical percolation transition in the Ising model. Techniques of using finite size scaling and fixed point behavior to search for the critical point are demonstrated.
Correlations of the differently ranked sizes of cluster, the maximum and second largest cluster ($C_{12}$), the maximum and third largest cluster ($C_{13}$), the second and third largest cluster ($C_{23}$), are studied in Section 3.
In Section 4,  we will investigate critical properties of skewness and kurtosis of the maximum, second and third largest cluster.  Signatures of  the fixed point, non-monotonic, and sign changes are systematically discussed, respectively.  The universal feature of fixed point in skewness and kurtosis is discussed in section 5. A possible application to relativistic heavy-ion collisions is also discussed. Finally, the results are summarized in Section 6.

\section{Percolation Transition in the Ising Model}

It has been pointed out that  the  QCD critical point belongs to the same universality class as 3D Ising model. In this case, magnetization $M$ is the order parameter. Here, $M=\frac{1}{V}\left<\sum\limits_{i}s_{i}\right>$, where $s_i$ is the spin at site $i$ which can take only two values $\pm1$. Using a parametric representation or  Monte Carlo simulations,  the Ising model can be extensively applied to study the QCD critical point.

Beside the research of the above thermal magnetization transition, it can also be applied to study the geometrical percolation transition. In this case,
 the spin $s_i$ at site $i$ can still take only two values $\pm1$. The thermal equilibrium configurations by using Wolff algorithm are still produced~\cite{wolff-algorithm}. It means that the system is already thermal when studying the geometrical percolation transition.  With a well-defined of the cluster, we can study both the thermal magnetization transition and geometrical percolation transition in the Ising model. In this paper,
the clusters are constructed by linking the nearest-neighbor spins with the same spin value. The size of a cluster ($n_s$) is the number of sites in a cluster. The maximum cluster size is corresponding to the largest $n_s$ in each event. Similarly, the values of the second and third largest cluster size are related to the second and third largest $n_s$, respectively.

The simulations are performed on different square lattices of size $N=L^2$ with $L = 50, 100, 150, 200$ without magnetic field. The Wolff cluster algorithm is employed to generate events~\cite{wolff-algorithm}.  With given system size and temperature, we totally run the program for 40,010,000 Wolff Monte Carlo steps, of which 2 million states are used in the calculations.  The initial 10,000 steps are used to insure equilibrium. To make sure the generated configurations are independent, we only use one state every 20 Monte Carlo steps. To treat the boundaries of the lattice,  various types of boundary conditions are employed in simulations, such as the periodic boundary condition, free boundary condition and  helical boundary condition and so on.
With different types of boundary condition, the scaling formula is still valid while the scaling function is different~\cite{Boundary}. The periodic boundary condition is employed in our simulations in this paper. The Bootstrap method is used to estimate the statistical uncertainty~\cite{Bootstrap, STAR-long-paper}. Except some results for kurtosis of the differently ranked sizes of the cluster, the statistical errors of other results are all smaller than the size of the symbols used in this paper.


\begin{figure}[th]
\centerline{\includegraphics[width=10.0cm]{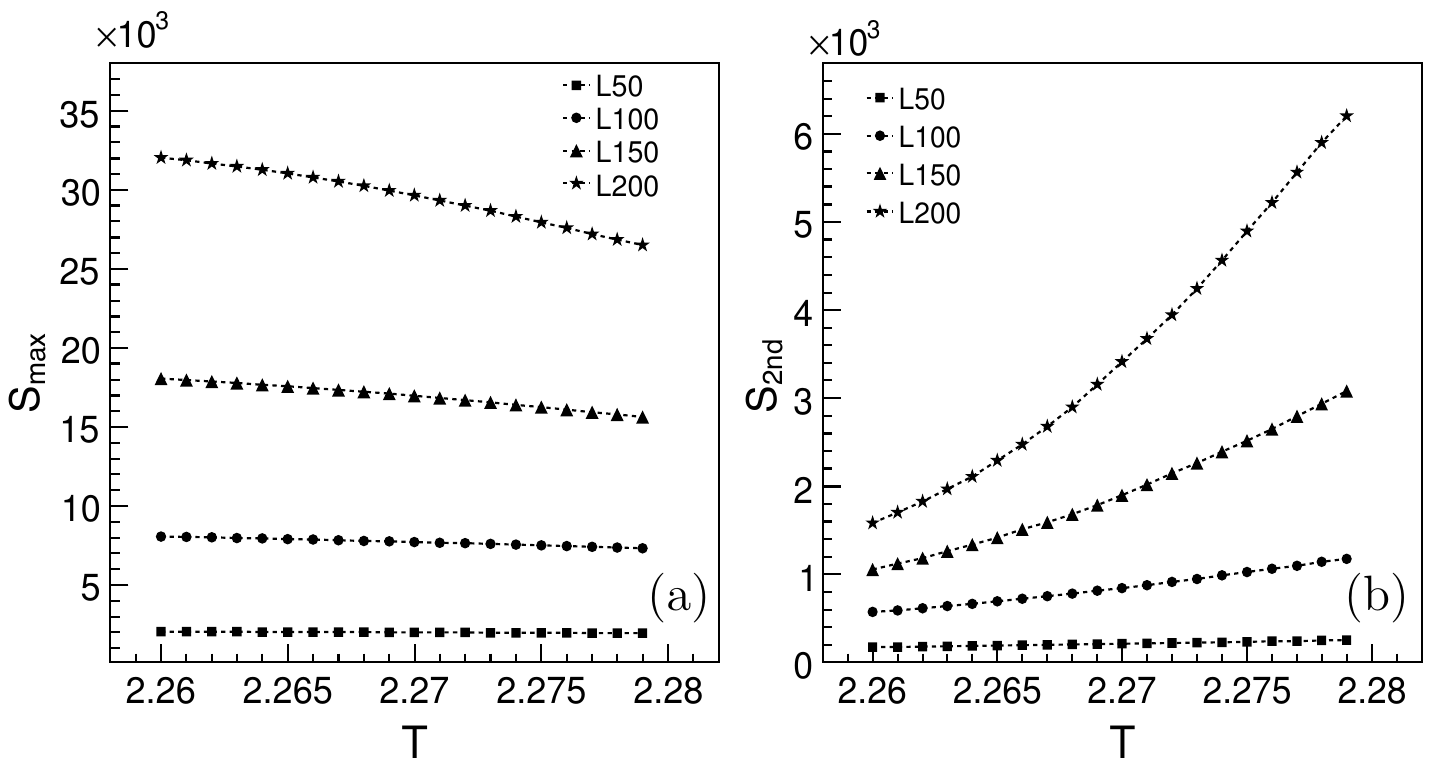}}
\caption{\label{distribution-first-second}(a) and (b) Distributions of the maximum and second
largest cluster size at different  system sizes near
the thermal critical point, respectively.}
\end{figure}

Figs.~\ref{distribution-first-second}(a) and (b) show the distribution of  the maximum and second largest cluster sizes near the thermal magnetization transition temperature, respectively. The size of the maximum largest cluster decreases with the temperature near the thermal magnetization critical point.  On the contrary, the size of the second largest cluster size increases with temperature.
It is hard to extract any information about critical point from these two original distributions. For the system with finite volume,   We should use the finite size scaling to study the critical behavior and search for the critical point.

According to the finite size scaling~\cite{Boundary, scaling-2}, the scaling function of the maximum cluster near the critical point is
\begin{equation}\label{12 scaling}
S_{\rm max}=L^{d-\beta_p/\nu_p} \cdot F_s(tL^{1/\nu_p}).
\end{equation}

Here $d=2$ is the spacial dimensionality  of the system. $t=\left(T-T_{c}\right)/T_{c}$ is the reduced temperature and $T_c$ is the critical temperature. $\nu_p$ and $\beta_p$ are the percolation exponents.  Eq.(\ref{12 scaling}) means that the scaled variable   $S_{\rm max}/L^{d-\beta_p/\nu_p}$ with different values of $t$ and $L$ can be described by a single scaling function $F_s(tL^{1/\nu_p})$.

\begin{figure}[th]
\centerline{\includegraphics[width=10.0cm]{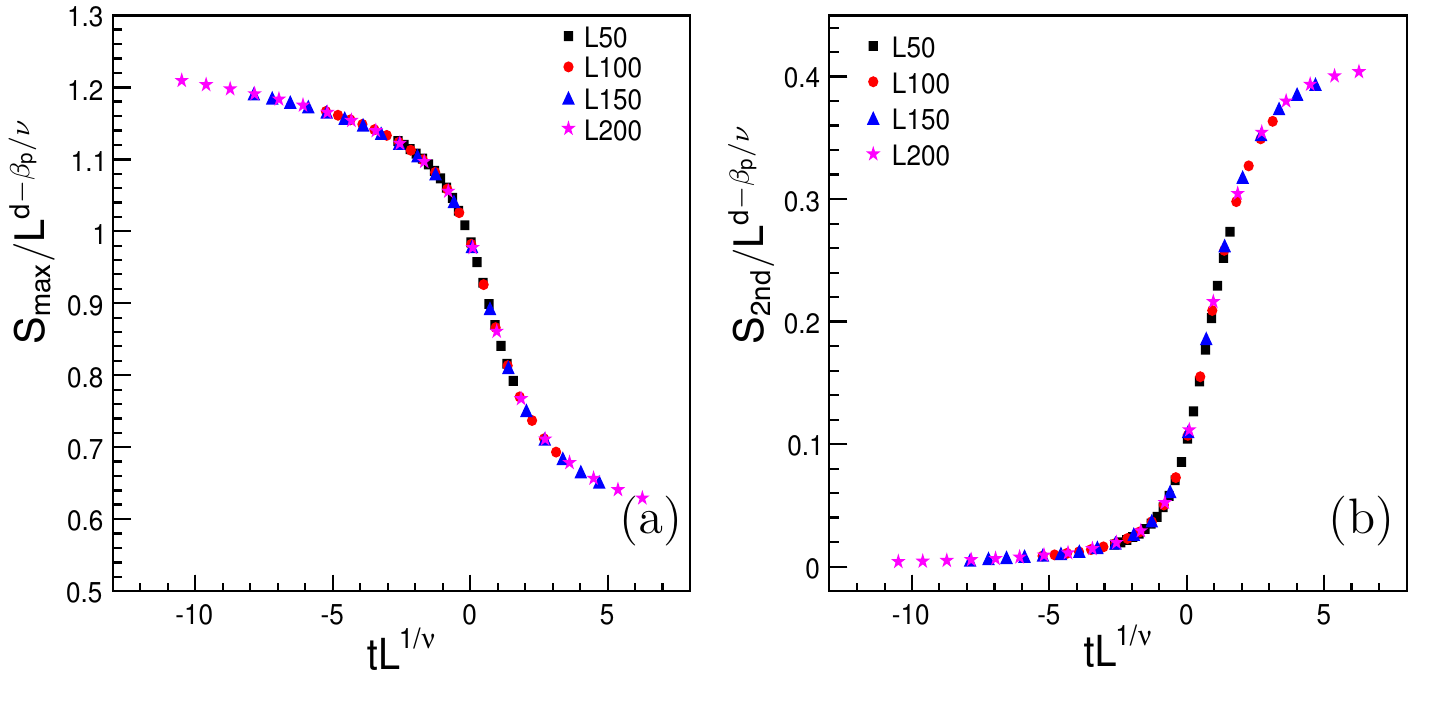}}
\caption{\label{scaling-first-second} (Color online) (a) and (b) The finite size scaling of
the maximum and second largest cluster, respectively.}
\end{figure}

Figs.~\ref{scaling-first-second}(a) and (b) demonstrate finite size scaling of the maximum  and second  largest cluster  near the critical point. The values of  $\beta_p$ and $\nu_p$ are consistent with critical exponents of the percolation transition~\cite{percolation-exponent}.
The exponent of the percolation correlation length $\nu_p$ is equal to the exponent of the magnetization correlation length $\nu_m$,  $\nu_p$ = $\nu_m$ =1.  And the critical percolation temperature coincides with the critical magnetization temperature.
It can be clearly observed that they both satisfy the finite
 size scaling in the whole critical region. Although the scaling function is different,  the critical exponent of the second largest cluster is the same as that of the maximum largest cluster.

Currently, the finite size scaling has been applied to study the QCD critical point in heavy-ion collisions~\cite{finite-size-Roy-1, finite-size-Roy-2}. However, as we had discussed in Ref.~[\refcite{Potts-2}], this method often lacks precision, since it is hard to select the best scaling function through visual observation.

An improved method to identify the critical point is the fixed point behavior~\cite{Potts-2}. At $T=T_c$,  the scaling function $F_s$ in Eq. (\ref{12 scaling}) becomes
\begin{equation}\label{12 fix-point}
F_s(0) = S_{\rm max}/L^{d-\beta_p/\nu}.
\end{equation}
It shows $S_{\rm max}/L^{d-\beta_p/\nu}$ is independent of system size $L$ at $T=T_c$. If we plot  $S_{\rm max}/L^{d-\beta_p/\nu}$ as a function of $T$, all curves of different system sizes intersect exactly at a fixed point, i.e., the critical point.

\begin{figure}[th]
\centerline{\includegraphics[width=10.0cm]{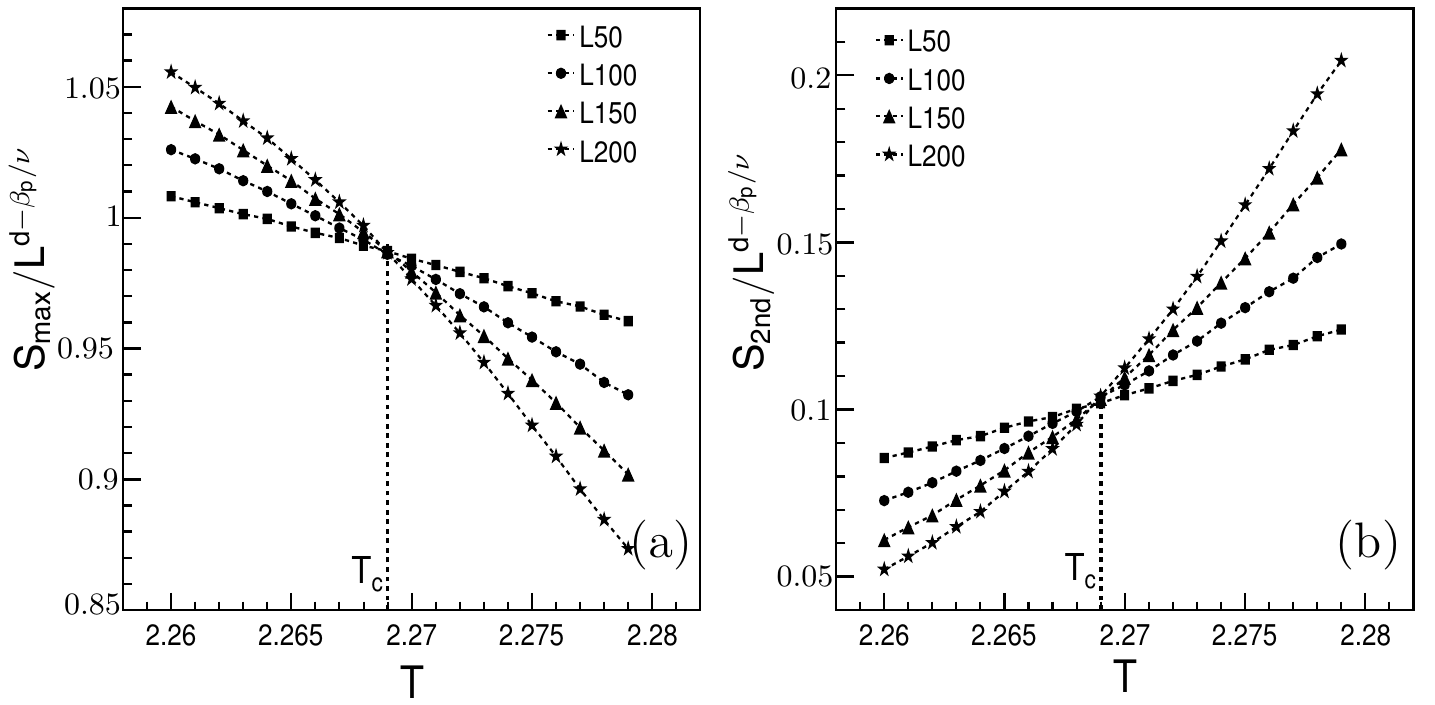}}
\caption{\label{fixed-point-first-second} (a) and (b) The fixed point extracted from
the maximum largest and second largest cluster, respectively.}
\end{figure}

Figs.~\ref{fixed-point-first-second}(a) and (b) present temperature dependence of the
 the scaled maximum largest cluster  $S_{\rm max}/L^{d-\beta_p/\nu}$ and second largest cluster   $S_{\rm 2nd}/L^{d-\beta_p/\nu}$, respectively. With different system sizes, the curves remarkably intersect at $T=T_c$. Here we want to address that $S_{\rm max}$ and $S_{\rm 2nd}$ should be scaled by $L^{d-\beta_p/\nu}$ to obtain the fixed-point. Mathematically, the goodness of the fixed point behavior can be quantified ~\cite{Potts-2}. However, it is still different with the observable discussed in Eq.~(\ref{FSS-fixed-directly-2}), which don't need to be re-scaled by other variable. The observables that can directly observe the fixed point will be demonstrated in the following.

\section{Correlations of the Differently Ranked Sizes of Cluster}

As demonstrated in Fig.~\ref{distribution-first-second},  in vicinity of the critical point,  the size of the maximum largest cluster is a decreasing function of temperature, while the size of the second largest cluster increases as a function of temperature.
It indicates the negative correlation of the maximum and the second largest cluster ($C_{12}$). The quantity of $C_{12}$ is defined as
\begin{equation}\label{correlation-definition}
C_{12}= \frac{\langle S_{\rm {max}} S_{\rm {2nd}}\rangle}{\langle
S_{\rm {max}}\rangle \langle S_{\rm {2nd}}\rangle}-1.
\end{equation}

\begin{figure}[tbp]
\centerline{\includegraphics[width=10.0cm]{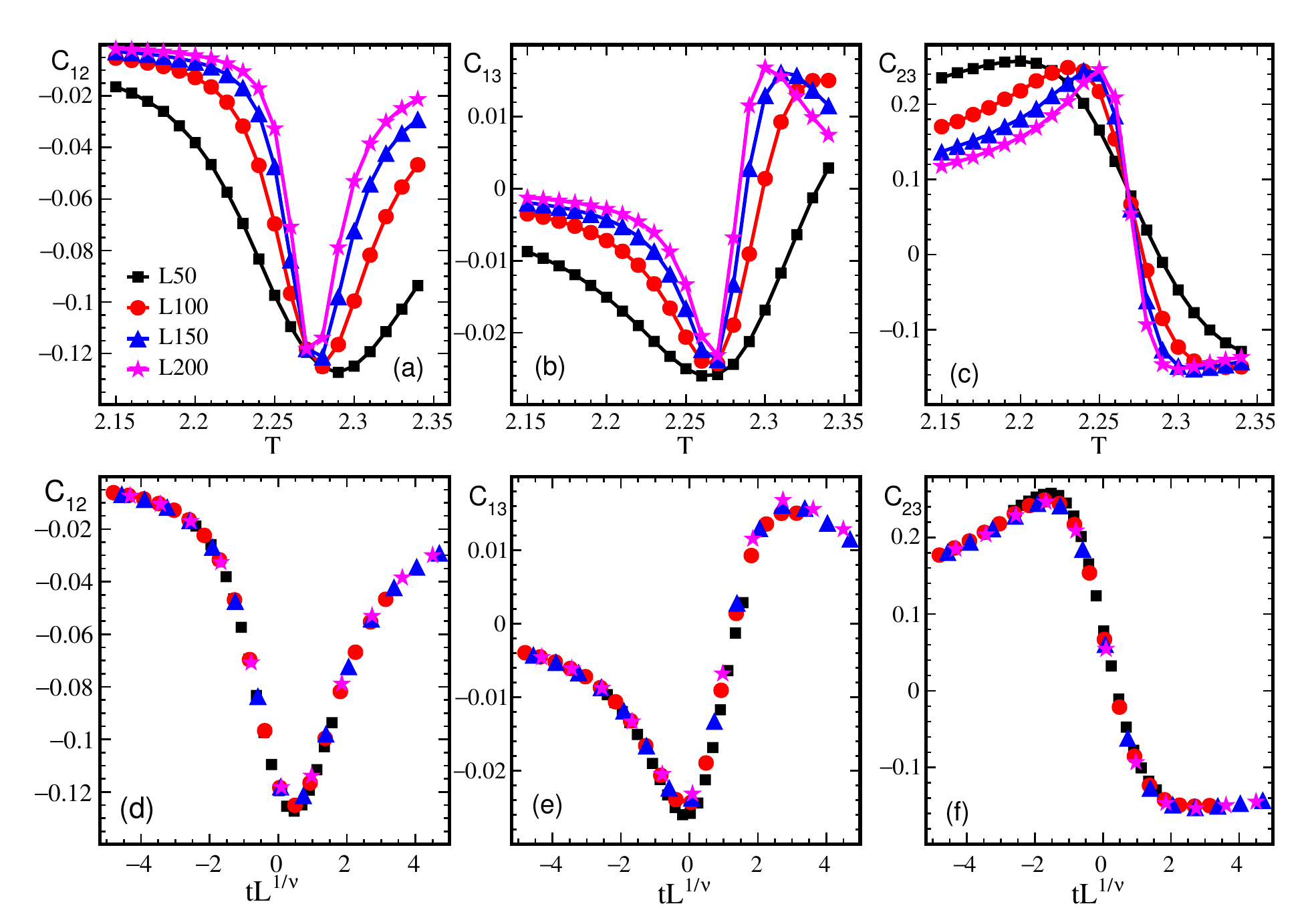}}
\caption{\label{correlation} (Color online) Upper panel: (a), (b) and (c) Temperature dependence of  $C_{12}$ , $C_{13}$ , and
$C_{23}$, respectively. The lines are just a guide to the eye. Lower panel: the same measurements as in corresponding upper panel, but they are as functions of the  scaling
variable $tL^{1/\nu}$.}
\end{figure}

Fig.~\ref{correlation}(a) shows $C_{12}$ as a function of $T$ in vicinity of the critical point.  With four different lattice sizes ($L=50, 100, 150, 200$), the  non-monotonic behavior is observed and the values of $C_{12}$ are all negative.
It first drops down  below $T_c$, and then goes up close to zero.  Figs.~\ref{correlation}(b) and (c) show that the  correlation of the maximum and the third largest cluster ($C_{13}$) and the correlation of the second and the third largest cluster ($C_{23}$) are not 0, either.  When $T>T_c$, $C_{13}$ significantly changes  from negative to positive. In contrary, $C_{23}$ shows a sharp change from positive to negative near the critical point.  In addition, the values of $C_{23}$ are one order of magnitude bigger than that of  $C_{12}$ and $C_{13}$.

The structure of those correlations are different. However, Figs.~\ref{correlation}(a) to (c)  show that with different system sizes, the fixed point behavior can be directly observed at  critical temperature. They do not need to be re-scaled by other variable. Theoretically, this feature is very helpful to search for the critical point. It is similar with the Binder-ratio, which has been extensively used to search for the critical point~\cite{Binder}.

Since the critical point can be identified directly, one can expect the following finite size scaling formula:
\begin{equation}\label{C-scaling}
C (T, L)= F_{C}(tL^{1/\nu})
\end{equation}
Figs.~\ref{correlation}(d) to (f) show the scaling behavior of $C_{12}$, $C_{13}$, and $C_{23}$, respectively. With different system lattice sizes, the finite size scaling is held for these three observables.  This scaling formula is totally the same as that in Eq.~(\ref{FSS-fixed-directly-1}), though the scaling functions are different. Consequently, correlations of the differently ranked size of cluster are all sensitive to the critical point. They can directly identify the critical point through the temperature dependency with various different system sizes.

\section{Skewness and Kurtosis of the Various Sizes of Cluster Near the Critical Point }
In heavy-ion collisions, due to the finite size system and finite evolution time, enhancement of the critical correlations and fluctuations is suppressed.  High-order cumulants are therefore proposed to study the QCD phase transition.  It has been proposed that ratio of the high-order cumulants are sensitive to the critical point. On the other hand, skewness ($S$) and kurtosis ($\kappa$) can also give us more information about the distribution. Skewness is a measurement of the symmetry of a distribution whereas
kurtosis measures the peakedness of the
distribution. Skewness and kurtosis with the maximum, second and third largest sizes of cluster can be expressed as
\begin{equation}\label{definition-skewness}
{\rm Skewness}\rm\_n=\frac{\langle(S_n-\langle S_{n}\rangle)^{3}\rangle}{\langle(S_{n}-\langle S_{n}\rangle)^{2}\rangle^{1.5}},
\end{equation}
and
\begin{equation}\label{definition-kurtosis}
{\rm Kurtosis}\rm\_n=\frac{\langle(S_{n}-\langle S_{n}\rangle)^{4}\rangle}{\langle(S_{n}-\langle S_{n}\rangle)^{2}\rangle^{2}} - 3.
\end{equation}

\begin{figure}[th]
\centerline{\includegraphics[width=9.0cm]{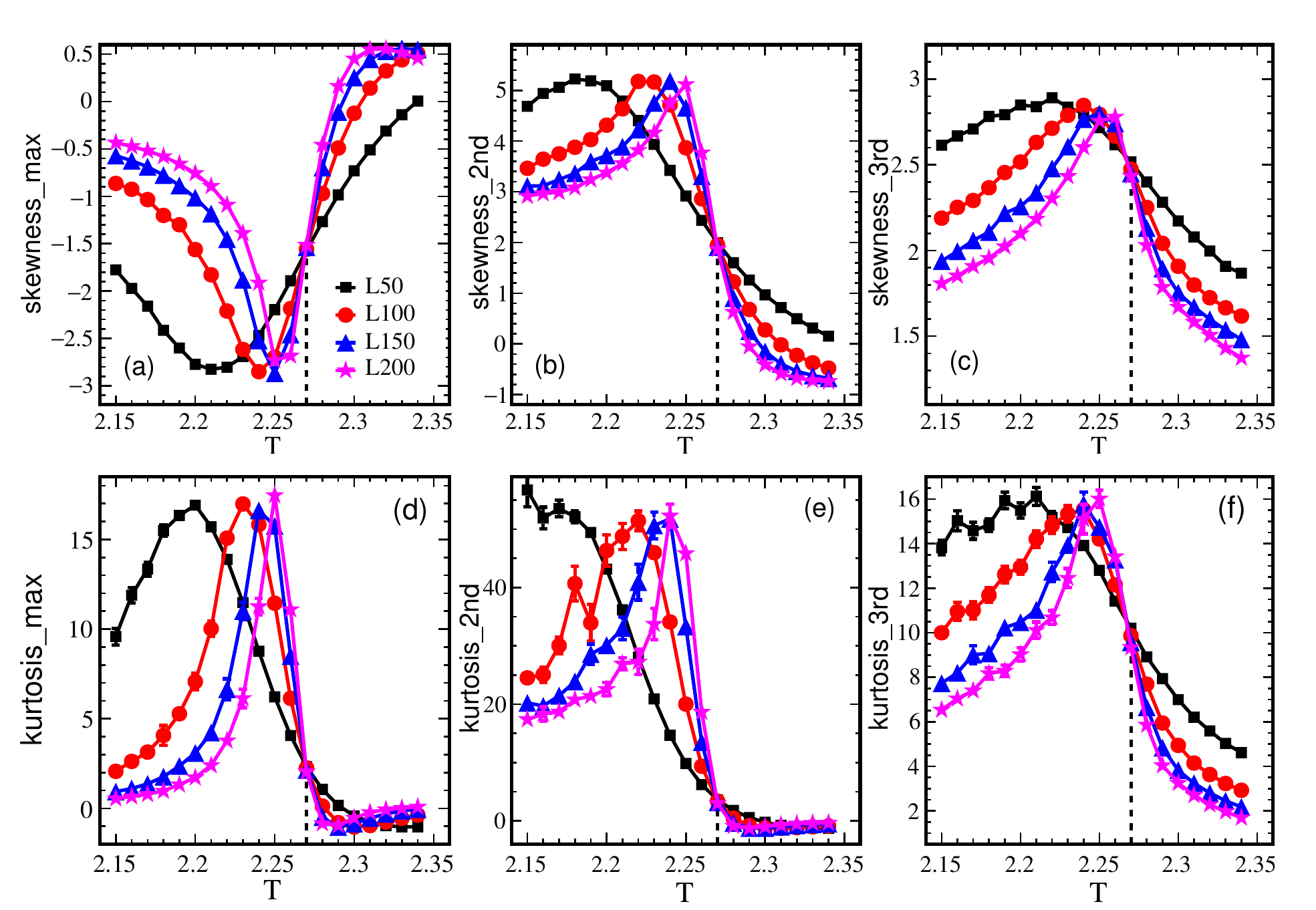}}
\caption{\label{plot-skewness} (Color online) (a) and (b) and (c) Temperature dependence of
skewness of sizes of the maximum cluster, the second largest cluster,
and the third largest cluster, respectively. (d) and (e) and (f) Temperature dependence of
kurtosis of sizes of the maximum cluster, the second largest cluster, and the third largest cluster, respectively.
The statistical uncertainty is obtained by Bootstrap method. The solid lines are just a guide to the eye.}
\end{figure}

Fig.~\ref{plot-skewness} demonstrates critical behavior of skewness and kurtosis of sizes of the maximum cluster ($\rm skewness_{\_max}$ and $\rm kurtosis_{\_max}$), the second largest cluster  ($\rm skewness_{\_2nd}$ and $\rm kurtosis_{\_2nd}$), the third largest cluster ($\rm skewness_{\_3rd}$ and $\rm kurtosis_{\_3rd}$), respectively. It is found that behavior of the sign change is not always robust, although it really exists. Fig.~\ref{plot-skewness}(a) shows negative  $\rm skewness{\_max}$ below $T_c$.  When $T>T_c$ it goes up and finally becomes positive. However, the positive signal is very small comparing to the negative phenomena. In addition, the positive value is hard to observe if the system size is not large enough, as shown by $L=50$.  When using the high-order cumulants to study the critical point, behavior of the sign change is not the necessary requirement.

Values of  $\rm skewness_{\_2nd}$ are systematically larger than  $\rm skewness_{\_max}$, and  $\rm kurtosis_{\_2nd}$ are also significantly larger than  $\rm kurtosis_{\_max}$. The previous section had shown that the values of $C_{23}$ are one order of magnitude bigger than that of  $C_{12}$ and $C_{13}$.  That is the reason why we propose studying the critical properties of the various sizes of cluster, instead only the maximum largest cluster.
On the other hand, Fig.~\ref{plot-skewness} also shows that values of the kurtosis are all one order of magnitude bigger than the corresponding skewness with differently ranked size of cluster. It means the higher order cumulant, the more sensitive to the critical point.  If the system size is large enough and the statistics are sufficient, we can study as many differently ranked clusters and high-order cumulants as possible.

 It is remarkable that the critical point can be directly identified from the temperature dependence of all these observables.
Values of the skewness and kurtosis are independent of the system size at $T=T_c$.  They are not zero within a broad range of temperature near the critical point.

Skewness and kurtosis of differently ranked sizes of cluster show colorful critical phenomena near the critical point. It is interesting to study these observables in CSPM or any other model that can give a connection between the percolation transition and heavy-ion collisions. Here we do not present critical behavior of ratios of the high-order cumulants of differently ranked sizes of cluster. As we will mention in the next section, they are also sensitive to the correlation length, but without the fixed point phenomena.

Finally,  the fixed point feature of skewness and kurtosis at the critical point is not the unique characteristic of the various sizes of cluster in 2D geometrical percolation transition.
The fixed point behavior of skewness and kurtosis can also be observed in thermal magnetization transition in the 3D Ising, $O(2)$, and $O(4)$ models~\cite{lizhu-cpc,On}.

\section{Universal Feature of Fixed Point in Skewness and Kurtosis}
With a finite size system, the critical behavior of the $n-th$ order cumulant ($K_n$) of order parameter can be described by the following formula:
 \begin{equation}\label{exponent-cumulant}
{K_n}=L^{nd}L^{-n\beta/\nu}F_{n}(tL^{1/\nu}).
\end{equation}
 This formula can be implemented to 2D and 3D thermal magnetization transition and geometrical percolation transition. Of course, the critical exponents $\beta$ and $\nu$ are different and determined by the universality class.

 Then the $n-th$ order susceptibility is measured by $K_n$ scaled by $L^d$. It means:
 \begin{equation}\label{scaling-susceptibility}
 \chi_n = L^{(n-1)d}L^{-n\beta/\nu}F_{n}(tL^{1/\nu}).
\end{equation}

For example, if we want to derive the scaling formula of the susceptibility:
\begin{equation}\label{scaling-2order}
{\chi_2}=L^{d-2\beta/\nu}F_{K_{2}}(tL^{1/\nu})=L^{\gamma/\nu}F_{2}(tL^{1/\nu}).
\end{equation}
Here, $\gamma = d\nu - 2\beta$ is derived from the following scaling relations of the critical exponents: $2-\alpha = d\nu$ together with $\alpha = 2-2\beta - \gamma$.


Eq.~(\ref{exponent-cumulant}) also tell us that the scaling forms of the skewness ($S=K_3/K_2^{3/2}$) and kurtosis ($\kappa=K_4/K_2^2$) are
 $S=F_{S}(tL^{1/\nu})$ and $\kappa=F_{\kappa}(tL^{1/\nu})$, respectively. They don't diverge with the system size.
 Values of skewness and kurtosis are finite and they are sensitive to the critical point with a broad range.
 At $T=T_c$, the value of skewness or kurtosis is a system size independent constant. The behavior of the system size independency at the critical point is independent of the dimensionality of the lattice or the number of dimensions of the order parameter. It is similar with the Binder-ratio, which is a powerful quantity to identify the critical point with the finite size system.
Consequently, the key point of application of the fixed-point feature is that we need to know the order parameter of the system. It can only be observed in skewness and kurtosis of the order parameter.

In heavy-ion collisions, the baselines of skewness and kurtosis of conserved quantities under many different distributions are investigated. In a normal distribution, the values of skewness and kurtosis are both zero. But beyond that, they are all determined by the parameters of the statistical distributions, such as the  Poisson distribution, the Skellam distribution, and the negative binomial distribution and so on~\cite{kurtosis-baseline}. Currently, the energy dependence of  $\kappa\sigma^2$ of net-proton number had already shown a hint of the non-monotonic behavior~\cite{STAR-non-monotonic}. However, the  measured skewness and kurtosis still follow general trends of those statistical distributions~\cite{cumulant-proton, cumulant-charge, cumulant-kaon}. It is interesting to study why $\kappa\sigma^2$ show the non-monotonic signal while $\kappa$ doesn't show the critical feature of a fixed point or a large deviation.
 As we know, there are still many uncertainties in applying the fixed point behavior to relativistic heavy-ion collisions, such as the finite time evolution, the volume fluctuation, and effect of the conserved law and so on. We will consider these effects in the future.


\section{Summary}

In summary, it is interesting to study critical properties of the various sizes of cluster in geometrical percolation transition model, which can help us to realize the formation of QGP. We suggest that we should firstly study percolation transition under finite size of the system, before applying to CSPM or any other model that can give a connection between the percolation transition and heavy-ion collisions.

In this paper, we study critical properties of various sizes of cluster with finite size of the system in 2D Ising model. We demonstrate the finite size scaling of the maximum and second largest cluster near the critical point. We also present how to use the fixed point behavior to search for the critical point.

We systematically study fluctuations and correlations of the differently ranked
sizes of the cluster, especially the quantities that we can directly identify the critical
point from the temperature dependency.
For the second largest cluster, the values of skewness and kurtosis are both one order of magnitude bigger than those extracted from the maximum largest cluster.  And the correlation $C_{23}$ is also one order of magnitude bigger than $C_{12}$ and $C_{13}$.  Consequently,  instead of only the maximum largest cluster, we should study as many  differently ranked sizes of cluster  as possible, if the statistics are sufficient and the system size is large enough.

Behavior of the sign change is not always significant for skewness and kurtosis of differently ranked sizes of cluster, although it really exists. The non-monotonic behavior is not obvious, either. It is not sufficient to draw the conclusion only based on one typical phenomenon or one observable. Instead, we should comprehensively study as many critical related observables as possible.

At critical point, the fixed point is observed from skewness and kurtosis with the maximum, second and third largest sizes of cluster, and correlations with the differently ranked sizes of cluster ($C_{12}$, $C_{13}$, and $C_{23}$). For skewness and kurtosis,
the fixed point behavior is universal and had already been observed in $O(1)$, $O(2)$ and $O(4)$ model with thermal dynamic transition. It is worth pointing out that this feature is similar with that derived from the Binder-ratio, which has provided a remarkable examination of the critical point. Currently, there are still many uncertainties when applying the fixed point behavior in relativistic heavy-ion  collisions.
 However, it is still interesting to study the reason why it is not observed in skewness and kurtosis in heavy-ion collisions.

\section*{Acknowledgements}

This work was supported by the postdoctral science and technology project of Hubei Province under Grant No. 2018Z27, and the Fundamental Research Funds for the Central Universities under Grant No. CCNU19ZN019.


\begin{thebibliography}{0}
\bibitem{crossover} Y. Aoki, G. Endrodi, Z. Fodor et al. Nature, {\bf 443}: 675 (2006).
\bibitem{first-order-1} S. Ejiri, Phys. Rev. D, {\bf 78}: 074507 (2008).
\bibitem{first-order-2} E. S. Bowman and J. I. Kapusta, Phys. Rev. C, {\bf 79}: 015202 (2009).
\bibitem{cumulant-proton} L. Adamczyk et al. (STAR Collaboration), Phys. Rev. Lett. {\bf 112}: 032302 (2014).
\bibitem{cumulant-charge} L. Adamczyk et al. (STAR Collaboration), Phys. Rev. Lett. {\bf 113}: 092301 (2014).
\bibitem{cumulant-kaon} L. Adamczyk et al. (STAR Collaboration), Phys. Lett. B {\bf 785}: 551 (2018).
\bibitem{qcd-transition-Karsch-1} A. Bazavov et al. (HotQCD Collaboration), Phys. Lett. B {\bf 795}: 15 (2019).
\bibitem{qcd-transition-Karsch-2} A. Bazavov et al. (HotQCD Collaboration), Phys. Rev. Lett. {\bf 123}: 062992 (2019).
\bibitem{qcd-transition-Fodor} Y. Aoki, S. Borsanyi, S. Durr et al. JHEP {\bf 06}: 088 (2009).
\bibitem{qcd-transition-Gupta} S. Gupta, X. Luo, B.Mohanty et al. Science {\bf 332}: 1525 (2011).
\bibitem{qcd-sign-problem} A. Bazavov et al. (HotQCD Collaboration), Phys. ReV. D {\bf 96}: 074501 (2017).
\bibitem{On} Xue Pan, Lizhu Chen, X. S. Chen et al. Nucl. Phys. A {\bf 913}: 206 (2013).
\bibitem{stephanov-1} M. A. Stephanov, Phys. Rev. Lett. {\bf 102}: 032301 (2009).
\bibitem{stephanov-2} M. A. Stephanov, Phys. Rev. Lett. {\bf 107}: 052301 (2011).

\bibitem{Ising-slowing-down-1} M. A. Stephanov, K. Rajagopal, and E. V. Shuryak, Phys. Rev. D, {\bf 60}: 114028 (1999).

\bibitem{Ising-slowing-down-2} B. Berdnikov and  K. Rajagopal, Phys. Rev. D, {\bf 61}: 105017 (2000).

\bibitem{finite-size-Roy-1} Roy A. Lacey, Phys. Rev. Lett, {\bf 114}: 142301 (2015).
\bibitem{finite-size-Roy-2} Roy A. Locey, Peifeng Liu, Niseem Magdy et al. arXiv:1606.08071 [nucl-ex].
\bibitem{STAR-non-monotonic} J. Adam et al. (STAR Collaboration), Phys. Rev. Lett. {\bf 126}: 092301 (2021).
\bibitem{percolation-QGP-1} T. Celik, F. Karsch, H. Satz, Phys. Lett. B {\bf 97}: 128 (1980).
\bibitem{percolation-xu} Xu Ming-Mei, Yu Mei-Ling, Liu Lian-Shou, Phys. Rev. Lett. {\bf 100}: 092301 (2008).
\bibitem{percolation-yu} Yu Mei-Ling, Xu Ming-Mei, Liu Zheng-You et al. Chin. Phys. C {\bf 33}: 552 (2009).
\bibitem{percolation-2d} Brijesh K Srivastava, Nucl. Phys. A {\bf 926}: 142 (2014).
\bibitem{percolation-2d-2} M. A. Braun, J. Dias de Deus, A. S. Hircsh et al. Phys. Rep. {\bf 599}: 1 (2015).
\bibitem{percolation-QGP-2} R. P. Scharenberg, B. K. Srivastava, and  C. Pajares, Phys. Rev. D {\bf 100}: 114040 (2019).
\bibitem{percolation-Stauffer} D. Stauffer, Phys. Rep. {\bf 54}: 1 (1979).
\bibitem{Ising-temperature} A. Coniglio, C. R. Nappi, F. Peruggi et al.  J. Phys. A {\bf 10}: 205 (1977).
\bibitem{percolation-cluster} D. Stauffer and A.Aharony, Introduction to Percolation Theory 2nd.
ed. (Taylor and Francis, London, 1992).
\bibitem{Binder} K. Binder, Z. Phys. B {\bf 43}, 119 (1981); K. Binder, Rep. Prog. Phys. {\bf 60}:  487 (1997).
\bibitem{wolff-algorithm}  U. Wolff, Phys. Rev. Lett. {\bf 62}: 361(1989).
\bibitem{Boundary} V. Privman, Finite-size Scaling and Numerical Simulation of Statistical Systems (World Scientific, Singapore, 1990) p35-37.
\bibitem{Bootstrap} B. Efron, Computers and the Theory of Statistics: Thinking the Unthinkable (Society for Industrial
 and Applied Mathematics, 1979).
\bibitem{STAR-long-paper} M. S. Abdallah et al. (STAR Collaboration), arXiv:2101.12413 [nucl-ex].
\bibitem{scaling-2} J. Cardy, Scaling and Renormalization in Statistical Physics (Cambridge University Press, Cambridge, UK, 1996).
\bibitem{percolation-exponent} A. L. Stella and C. Vanderzande, Phys. Rev. Lett. {\bf 62}:
1067(1989).
\bibitem{Potts-2} Yanhua Zhang, Yeyin Zhao, Lizhu Chen et al. Phys. Rev. E {\bf 100}: 052146 (2019).
\bibitem{lizhu-cpc} Chen Li-Zhu, Pan Xue, Chen Xiao-Song et al. Chin. Phys. C {36}: 727 (2012).
\bibitem{kurtosis-baseline} X. Luo, B. Mohanty, and N. Xu, Nucl. Phys. A {\bf 931}: 808 (2014).


\end{thebibliography}
\end{document}